\documentclass[a4paper,10pt,twocolumn,3p]{elsarticle}
\biboptions{sort&compress}
\journal{ }
\usepackage{amsmath}
\usepackage{amsfonts}
\usepackage{hyperref}
\PassOptionsToPackage{linktocpage}{hyperref}
\usepackage{dcolumn}     
\usepackage{bm}          
\usepackage{graphicx}
\usepackage{epsfig}
\usepackage{epstopdf}
\usepackage{xcolor}

\usepackage{soul}
\renewcommand*{\today}{ }

\usepackage{etoolbox}
\makeatletter
\patchcmd{\ps@pprintTitle}
  {Preprint submitted to}
  {\textcopyright \, 2018. This is made available under the CC-BY-NC-ND 4.0 license, \linebreak http://creativecommons.org/licenses/by-nc-nd/4.0/ .}
  {}{}
\makeatother

\begin{document}
\begin{frontmatter}


\title{A simple descriptor for energetics at fcc-bcc metal interfaces\tnoteref{t1}}
\tnotetext[t1]{Manuscript accepted on $10^\textrm{th}$ January $2018$ and published as `Linda A. Zotti, Stefano Sanvito, and David D. O'Regan, Materials and Design \textbf{142} (2018) 158Ð165', for which see \href{https://doi.org/10.1016/j.matdes.2018.01.019}{https://doi.org/10.1016/j.matdes.2018.01.019}}
\author{Linda A. Zotti$^{1,2}$}
\ead{linda.zotti@uam.es}

\author{Stefano Sanvito$^{2}$}
\author{David D. O'Regan$^{2}$}

\address{$^{1}$Departamento de F\'{\i}sica 
Teorica de la Materia Condensada, Universidad Aut\'onoma de Madrid, 28049 Madrid, Spain}
\address{$^{2}$School of Physics, AMBER and CRANN Institute, Trinity College Dublin, Dublin 2, Ireland}
\date{\today}


\begin{abstract}
We have developed a new and user-friendly 
interface energy calculation method that  avoids problems deriving
 from numerical differences between bulk and slab calculations, such as the number of $k$ points along the 
direction perpendicular to the interface. We have applied this  to 36 bcc-fcc metal interfaces in the (100)
 orientation and found a clear dependence of the interface energy on the difference between the work
 functions of the two metals, on the one hand, and the total number of $d$ electrons
on the other. Greater mechanical deformations were observed in  fcc crystals than in their bcc counterparts.
For each bcc metal, the interface energy was found to follow the position of its $d$ band, whereas
the same was not observed for  fcc.

\end{abstract}

\end{frontmatter}

\section{Introduction}
The study of metal-metal interfaces is crucial for many industrial processes and technological
 applications~\cite{hung2002,JIANG2017185,HOSEINIATHAR2015516}, including growth modes in thin films~\cite{raeker1994,sonderegger2009}, 
catalysis~\cite{palotas2016}, as well as many experimental techniques used in nanotechnology 
such as those involving metallic tips  on metal surfaces~\cite{feldbauer2015}. Theoretical support
 in designing metallic interfaces is essential as
it can provide information that is  extremely difficult to extract  experimentally.
An example is given by  interface energies, which determine the nucleation barrier
 and the shapes of precipitates~\cite{lu2014,jung2010,sawada2013}, besides the stability and reliability of the whole
system.  These energies are  not directly accessible experimentally. 
 Thorough studies have been performed at the level of first-principles calculations on selected solid-solid
 interfaces, focusing on various aspects such as the film thickness~\cite{raeker1994},
  orientation~\cite{feldbauer2015},  magnetoresistance~\cite{haney2007},  magnetic anisotropy~\cite{Guan2016}
 ferromagnetic moments~\cite{fu1985}, as well as electronic~\cite{linghu2016,park2012,li2015,gonzalez2016}, mechanical~\cite{wang2006,ziaei_2015,WENG20161,JUNG201528}, and thermodynamic ~\cite{martin2012,martins2012,nazir2015,Ruihuan2016,HU2017175} properties.
Notwithstanding the detailed nature of these analyses,  they were quite often mostly focused on a very few  materials.
What is currently still missing is, for instance, a systematic analysis and a rule of thumb as to how
 to ``cherry pick'' materials and match them, ensuring stability of their interface at the same time.  
In this  work, therefore, we chose to follow a different approach. We  focused on only one crystal
 orientation (100) and on one type of relative dislocation between the two metals, but we 
 performed a systematic analysis spanning over 36 interfaces. These were  obtained by combining 6
 face-centered-cubic (fcc) crystals (Au, Ag, Cu, Ni, Pd, Pt)  with 6 body-centered-cubic
 (bcc) crystals (Cr, Mo, W, Nb, V, Ta). Such an approach has allowed us to formulate a descriptor for
 interface energies based on very similar conditions for all systems.
 
 In particular, we aimed 
to understand whether it is possible to predict trends in interface energetics on the basis of simple
 bulk properties. In previous work, on the basis of non-first-principles calculations~\cite{gautier1991},
 interface energies were found to depend on the balance between several quantities, including
 the number of $d$ electrons per atom in the interface layers and in the bulk, the bandwidth of the
 interface layers and the bulk, the cohesive energies, the Fermi levels and the intra-atomic potentials. 
We  show here, instead, that modern density functional theory (DFT) calculations make it possible to reveal
 much simpler relationships.
In particular, we  show that, for certain metal pairs,  simply  the difference between their work functions or the sum of the electrons in their $d$ bands can provide a first hint 
on the stability of the interface.
We anticipate that this will prove to be a very useful finding for the design
of metallic multi-layers and heterostructures for technological applications.
Our work is particularly timely and relevant for interface layer selection and 
design in the context of high-throughput materials simulation and informatics, a research area which is gaining increasing traction at present.

\section{Methods}
The interface energy is the energy cost associated with the introduction of an interface. It can
be interpreted as the surface ``binding'' energy density of the two components.
It comprises two contributions, namely the chemical and  electronic energy
that originates from breaking and creating bonds to form an interface, and the elastic energy required
to create the interface by matching the two lattices~\cite{fors2010}.
We  mainly focus on the electronic component and neglect the elastic contribution, which goes beyond the 
scope of the present work.
Within one of the most accurate methods to date~\cite{lu2013}, the interface energy $\gamma$ can be calculated as:
\begin{equation}
 \gamma = E'_\textrm{fcc/bcc} -E_\textrm{fcc}^\textrm{bulk}- E_\textrm{bcc}^\textrm{bulk}, 
\label{secondmethod}
\end{equation}
where
\begin{equation}
 E'_\textrm{fcc/bcc}= E_\textrm{fcc/bcc} - \sigma_\textrm{fcc} -\sigma_\textrm{bcc}.
\label{sec-meth2}
\end{equation}

Here, $E_\textrm{fcc/bcc}$ is the total energy of the system, $E_\textrm{x}^\textrm{bulk}$ is the total energy of the crystal in the bulk state.  
The surface energies $\sigma_\textrm{x}$ are calculated relative to the bulk crystal experiencing the same strain
 as in the interface (see Ref.~\cite{lu2013} for  details), via
\begin{equation}
\sigma_\textrm{x}=( E_\textrm{x}-E_\textrm{x}^\textrm{bulk})/2.
\end{equation}
Although this method has proven successful~\cite{lu2013}, it requires five separate calculations (one for the interface, two for the corresponding bulk materials and two for their free surfaces), which  
require particular care when defining the unit-cell dimensions and the corresponding strain in each of them.
More importantly, it carries the intrinsic problem of performing algebraic operations between quantities
 that are calculated using different numbers of $k$ points. In fact, the lattice periodicity is preserved in all
 three space directions in bulk crystals, but only in two directions at surfaces. Consequently, energies computed for bulk
 and slab calculations are not directly comparable for all thickness values and  slow convergence of
 the calculated interface energy with respect to the number of layers can thereby arise. Slight variations of Eq.~\ref{secondmethod}
have also been used~\cite{zhao2010}, but these present the same problem. 

We thus propose an alternative method, namely
an extension of the Fiorentini procedure~\cite{fiorentini1996}  that was originally
 developed to calculate surface energies,  and which has been  shown to provide accelerated convergence with respect
to the number of layers in the system~\cite{singh2009}.
Surface energy is the energy needed to cleave a bulk crystal into two separate surfaces~\cite{raeker1994}
and it can be expressed as 
\begin{equation}
 \sigma=\lim_{N \to \infty}\frac{1}{2}({E^\textrm{slab}_{N}-NE^\textrm{bulk}}), 
\label{eq-fiorentini}
\end{equation}
where E$^\textrm{slab}_{N}$ is the total energy of an $N$-atom slab and E$_\textrm{bulk}$ is the total energy of the bulk per atom.
Within the Fiorentini method, E$_\textrm{bulk}$ can be calculated as the slope in E$^\textrm{slab}_{N}$ plotted against $N$ and  then used in equation~\ref{eq-fiorentini}. This is possible because the following linear relationship applies: 
\begin{equation}
E^\textrm{slab}_{N} \approx 2 \sigma + N E^\textrm{bulk}.
\label{eq-N}
\end{equation}
This method allows us to avoid problems deriving from calculating E$^\textrm{slab}_{N}$
and E$^\textrm{bulk}$ with a different number of $k$ points. We note that, if one-atom unit cells are considered, $N$  equals the number of layers.

\begin{figure}[t]
\begin{center}
\includegraphics[width=\columnwidth]{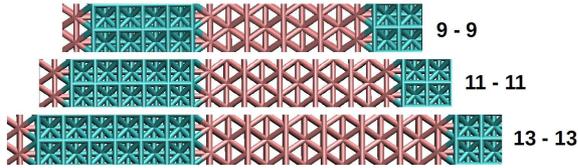}
\caption{Examples of slabs with different numbers of layers  in both fcc and bcc metals (indicated on the right hand side)
 as used in our extended ``Fiorentini'' approach for interfaces.} \label{slabs}
\end{center}
\end{figure}

By replacing the vacuum region with another metal, we have extended  this scheme  to the calculation
 of interface energies. In fact, in the same way that interface energies are the energies originating from
breaking old bonds and creating new bonds in the interface, surface energies can likewise be interpreted
as the energy involved in breaking the same old bonds and creating new ``bonds'' with the vacuum. 
Within such a scheme, we built, for each fcc-bcc interface, at least three structures which differ
from each other by the number of layers on each side (see Fig.~\ref{slabs}). 
After collecting the total energies of each structure, 
we extracted the slope $s$ of the total energy of the fcc-bcc system $E_{N_\textrm{x}+N_\textrm{y}}$ versus the
 total number of layers N$_\textrm{x}$+N$_\textrm{y}$, where N$_\textrm{x}$ and N$_\textrm{y}$ are the number of layers on each side of the interface.
The interface energy $\gamma$ is then given by
\begin{equation}
\gamma = (E_{N_\textrm{x}+N_\textrm{y}}-(N_\textrm{x}+N_\textrm{y})*s)/2, 
\label{eq-interf-fiorentini}
\end{equation}
where $E_{N_{x}+N_{y}}$ and $N_{x}+N_{y}$ must be taken from the same structure.
As for surface energies, this method avoids
problems arising due to numerical differences between bulk and slab calculations for interface energies.
In addition, it only requires two simple calculations (although it is recommendable to perform at 
least one more to
make sure the slope is evaluated in the linear regime). 
The work of separation $W$ (the energy needed to separate the interface
into two free surfaces) is given by~\cite{allan1974, hung2002}
\begin{equation}
 W = \sigma_\textrm{fcc}+\sigma_\textrm{bcc} - \gamma .
\label{eq-adhesion}
\end{equation} 

In order to calculate the necessary total energies, we carried out density functional theory (DFT) calculations by using the PWscf code of the
Quantum ESPRESSO distribution~\cite{QE}, which uses a plane-wave basis.
The LDA exchange-correlation functional was used, with a kinetic energy cutoff for wave-functions of 50 Ry
 and, for the charge density and potential, of 400 Ry. This 
functional was chosen because it has been found to yield better agreement with experiments for surface energies as
compared to GGA~\cite{singh2009} and because it is known to provide a very good description of structural
and energetic properties of solids~\cite{perez2000}. For the
 Brillouin-zone integration, we use a Monkhorst-Pack~\cite{Monkhorst} set. Specifically, we used a 16x16x16
 $k$-point grid for bulk calculations,
 while we used a 16x16x1 sampling for surface calculations.
A Fermi-Dirac smearing   with a broadening of 0.0038 Ryd ($\approx$ 600 K) was adopted. Ultrasoft
 pseudopotentials USPPs~\cite{pseudopotentials} were used for all elements. Although some of the systems analyzed are known to present interesting magnetic properties (for example, see Ref.~\cite{das2015}), 
 we forced all systems to be
 non-magnetic in order to keep their conditions as similar as possible and thus to better isolate the over-arching 
 descriptors for adhesion.

We considered fcc(100)-bcc(100) interfaces consisting of the same number $N$ of layers  in each metal.
Periodicity was applied in all three spatial directions. In order to meet this condition, odd values were
chosen for $N$, namely $N=9,11,13$. For each metal, a slab in contact with a vacuum region of approximately 16~{\AA}
 on each side was also built in order to calculate surface energies (for this kind of 
calculation, it does not matter whether an even or odd number of layers is used). 
To calculate surface energies, the size of the simulation cell was determined using optimized bulk lattice parameters and 
 all atomic coordinates were then relaxed. For the interface
 energies, instead,  cell dimensions were optimized  in all three Cartesian directions to
  ensure a minimal strain induced by the lattice mismatch at the interface.

We emphasise that our surface energies and interface energies were calculated using Eqs.~{\ref{eq-fiorentini}}-{\ref{eq-N}} and~{\ref{eq-interf-fiorentini}}, respectively. For selected cases, Eqs.~{\ref{secondmethod}}-{\ref{sec-meth2}} were also used for the purposes of comparison (see Table 2).

The (100) faces of the fcc and bcc metal were rotated
 by 45 degrees with respect to each other in order to provide  an optimal match between the bcc lattice 
constant a$_\textrm{bcc}$ and half a diagonal of the fcc face (a$_\textrm{fcc}/\sqrt{2}$, a$_\textrm{fcc}$ being the 
fcc lattice constant)~\cite{haney2007}.
We considered one-atom unit cells, suppressing, therefore, reconstruction effects and long-ranged mismatch.
In the ${100}$ orientation, both the bcc and fcc metals present an $abab$ layer stacking. 
 At the interface, the two materials were joined so that the atoms of one metal rested on top of the hollow
sites of the other. 

\section{Results and Discussion}
In the colour-coded areas of panel (a) and (b) of Table~\ref{interface}, we report the interface and
 work of separation calculated by  using Eqs.~\ref{eq-interf-fiorentini}
 and \ref{eq-adhesion}, respectively.
The ``Vacuum'' row and column in panel (a) refer to each individual metal interfaced with vacuum: this is nothing but 
the surface energy as pointed out above (note that, for Au, the calculated value is in very good agreement with
 that reported  in~\cite{singh2009}, where the same code and functional were used).  All corresponding values
 in J/nm$^{2}$ are shown in  Table S1 of the Supporting Information.
We can observe that the interface energies shown in panel (a) range from negative to positive values, which  
corresponds to more and less 
stable interfaces, respectively.

%
\begin{table}
\begin{center}
\includegraphics[width=\columnwidth]{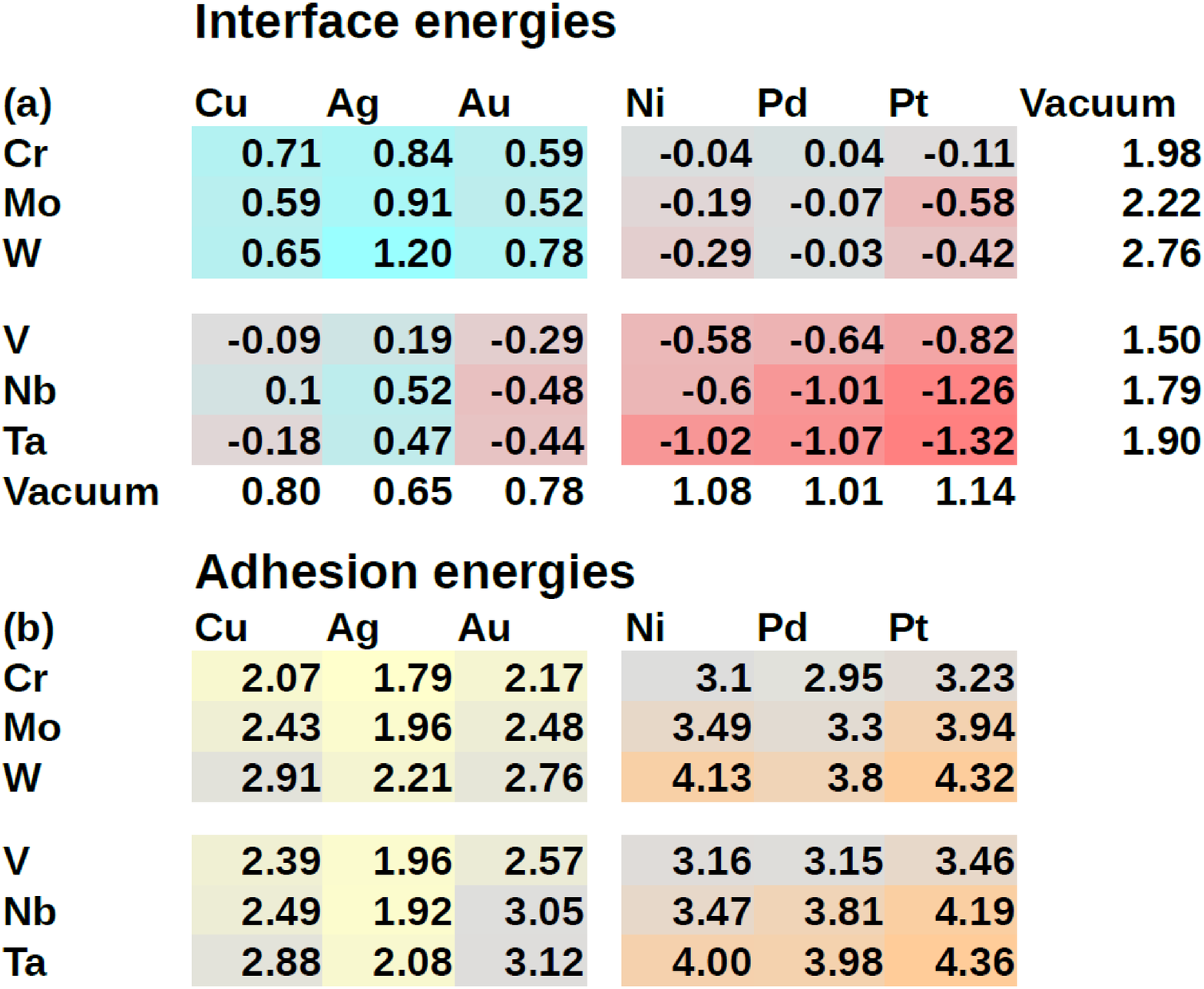}
\caption{ (a) Interface energies $\gamma $ and (b) Work of separation $W$. All values are given in eV/atom.} \label{interface}
\end{center}
\end{table}
%

Before analyzing these results in depth, it is worth comparing some of our values with those that we would
 obtain by the method described in Eq.~\ref{secondmethod} and  in Ref.~\cite{lu2013}. For this purpose,
 we chose one interface with a positive
(Ag-W) and one with a negative (Au-Nb) interface energy.  
In order to apply this method, the fcc-bcc interface was embedded in vacuum. All fcc (bcc) slabs
 contained an even number of layers, 12, so that the $abab$ stacking was preserved in the corresponding bulk
 calculations, in which  the fcc (bcc) block is repeated periodically in all three spatial directions.
In the bulk and surface calculations, the same lateral strain
 as results in the interface calculation was imposed.
The same exchange-correlation functional, pseudopotentials and parameters were used as in the calculations
performed according to Eq. {\ref{eq-interf-fiorentini}} and reported in Table~{\ref{interface}}. 
All calculated quantities used in the comparison for each system  are reported in Table {\ref{tbl:1}}. 
An interface energy difference of
0.03 eV and 0.11 eV was calculated for W-Ag  and Nb-Au, respectively, a  discrepancy which should ultimately
vanish in the limit of large layer number $N$.

\begin{table*}
\centering
\resizebox{\textwidth}{!}{%
\begin{tabular}{ | c | c | c | c | c | c | c | c | c | c | c | c | c | c | }
\hline
& \multicolumn{7}{c |}{Method of Ref.~\cite{lu2013}} & \multicolumn{6}{c |}{Proposed method}\\
\hline
 & bulk$_\textbf{fcc}$ & bulk$_\textbf{bcc}$ & surface$_\textbf{fcc}$ & surface$_\textbf{bcc}$ & $E_\textbf{fcc/bcc}$ & $\gamma$ (Ry) & $\bm{\gamma}$ \textbf{(eV)} & E(9-9) & E(11-11) & E(13-13) & $s$ & $\gamma$ (Ry) &$\bm{\gamma}$ \textbf{(eV)} \\
\hline
 & & & & & & & & & & & & & \\
{\bf Ag-W} & -3548.50 & -1910.83 & -3548.38 & -1910.43 & -5458.98 & 0.09 & \bf{1.23} &-4094.32 &-5004.21 &-5914.10 &-227.47 &0.09 &\bf{1.20} \\
& & & & & & & & & & & & & \\
{\bf Au-Nb}  & -1378.99 & -1417.12 & -1378.84 & -1416.86 & -2795.93 & -0.03 & \bf{-0.37} &-2097.14 &-2563.16 &-3029.17 &-116.50 &-0.04&\bf{-0.48} \\
& & & & & & & & & & & & &\\
\hline
\end{tabular}}
\caption{Interface energies calculated, for selected interfaces, using the method described in Ref.~{\cite{lu2013}} 
and  that which we propose, together with all values of the quantities employed in both. 
Where not specified, values are expressed in Rydberg. $E_\textrm{fcc/bcc}$ is the energy 
of the fcc-bcc interface embedded in vacuum,  bulk$_\textrm{x}$ is the energy of the two bulk metals,  surface$_\textrm{x}$
is the energy of the free surface; $E(N_\textrm{fcc} -N _\textrm{bcc}$) is the energy of a system consisting of an $N$-layer fcc slab
 and an $N$-layer bcc slab as shown in Fig.~{\ref{slabs}},  $s$ is the slope of $N_\textrm{fcc} - N_\textrm{bcc}$ versus $N_\textrm{fcc} +N_\textrm{bcc}$, and 
 $\gamma$ is the interface energy.}
\label{tbl:1}
\end{table*}

We furthermore calculated the interface energy for a Ag-Fe system in the same
orientation and found a value (1.07~J/m$^{2}$) that is in very good agreement with those reported for the same
interface in Ref.~\cite{lu2013}. 
Although the two methods give comparable results when the recipe described in Eq.~\ref{secondmethod} is carefully applied,
 the procedure proposed here is 
easier to follow and requires fewer types of calculations. 
It can also be used
for interfaces involving materials with different crystal structures, provided that care is taken when defining
the number of atoms per unit cell and the number of layers (chosen to enforce periodicity at the boundaries).  
 
Each of Table~\ref{interface} (a) and (b) is divided into four  blocks in order to highlight elements coming from different rows of the periodic table.
 Specifically, V, Nb and Ta belong to the 3$^\textrm{rd}$  column
of the transition-metal block; Cr, Mo and W belong to 4$^\textrm{th}$  column of the same block; 
Ni, Pd and Pt belong to the 8$^\textrm{th}$  column; and Cu, Ag and Au belong to the 9$^\textrm{th}$ column. 
These specific columns were chosen because, within each of them, the crystal structure is the same across three 
consecutive rows of the periodic table.
 The same does not apply instead, for instance, to the 5$^\textrm{th}$ to 7$^\textrm{th}$ columns, where a mixture of fcc, bcc and hcp structures is present.
We anticipate that our qualitative findings will extend to other metal-metal interfaces structures across the periodic table, provided that lattice mismatch, long-range reorganisation, magnetism, or more exotic many-body effects are absent or negligible. 

To make it easier for the reader to identify interfaces with similar values of interface energy and work of separation, we varied the cell colours in panel (a) from red (lowest negative values) to
blue (highest positive values) and in panel (b) from yellow (lowest positive) to orange (highest positive). 
The colour-coded areas in panel (a) highlight a tendency towards 
positive interface energies
for the 4$^\textrm{th}$-9$^\textrm{th}$ column combinations, and towards negative interface energies
for the 3$^\textrm{rd}$-8$^\textrm{th}$ column combinations. The sign of the interface energy plays an important role in
 inter-diffusion and interface stability~\cite{roder1993}. Negative values indicate a greater
stability of the interface. For the work of separation $W$,
 all calculated values are positive but the
color-coded areas show a similar color modulation as for the interface energies. 
To get an insight into the variation observed for these quantities, we investigated
possible dependencies on bulk properties, 
 with the aim of developing an approach to optimise
 adhesion based only on inputs from bulk (high-throughput compatible) calculations. 
 
\begin{figure}[t]
\begin{center}
\includegraphics[width=\columnwidth]{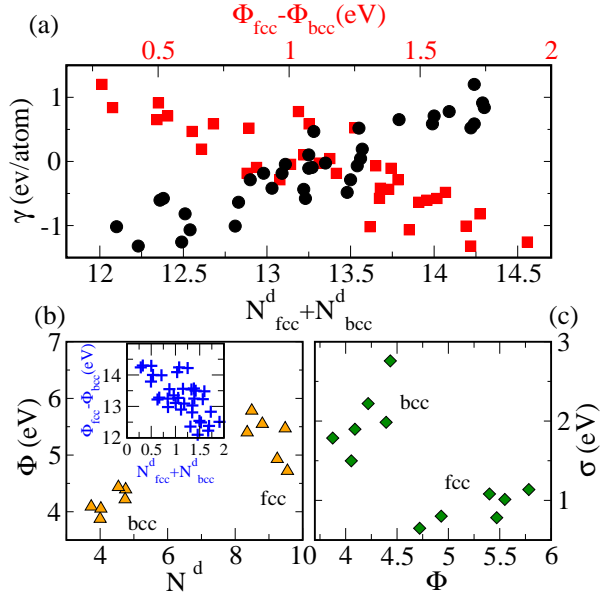}
\caption{(a) interface energy as a function of the sum of the total number of $d$ electrons $N^{d}$ (black dots) and as a function
 of the difference between the bulk  Fermi energies of the fcc and bcc metal (red squares); (b) work function as a function of $N^{d}$ and (inset) difference of work functions versus sum of $N^{d}$ in the two metals; (c) surface energies as a function of work function.} \label{trends-occup}
\end{center}
\end{figure}

In panel (a) of Fig.~\ref{trends-occup}, we show the interface energies against (black dots)
the sum of  the $d$-band occupation $N^{d}$ in the two metals (we focussed on this angular momentum because it is known to
play a dominant role in binding strength for transition metals \cite{norskov2011,tang2009}). 
 This occupation was evaluated using a separate bulk unit-cell calculation,  
 integrating the $d$ band density of states up to the Fermi energy. A linear trend can be observed,
showing negative energies for lower values of the sum and positive in the opposite case.
In the same panel we also report (red squares)  the interface-energy values as a function of the difference between
 the work functions of the two elements forming the junction. 
Work functions were calculated as
\begin{equation}
 \Phi = V_\textrm{Vacuum} - E_\textrm{Fermi}, 
\label{work function}
\end{equation}
where V$_\textrm{Vacuum}$ is the vacuum potential and $E_\textrm{Fermi}$ is the Fermi energy 
(see Fig.~S1 in the Supporting Information).
Both quantities were extracted from
calculations performed on the same slabs that were used to compute  surface energies. For the metals studied
in this work, work functions were generally found to be larger for fcc metals than for bcc ones (see Table~\ref{tbl:3} for values). 

\begin{table}[h]
\centering
\begin{tabular}{ | c | c | c | }
\hline
& Work function (eV) & $N^{d}$ \\
\hline
 Ag& 4.720 & 9.54717\\  
 Au& 5.470 & 9.48133\\
 Cr& 4.394 & 4.75568\\
 Cu& 4.930 & 9.24695\\
 Mo& 4.219 & 4.73858\\
 Nb& 3.873 & 4.00027\\
 Ni& 5.399 & 8.35574\\
 Pd& 5.549 & 8.8055 \\
 Pt& 5.783 & 8.48978\\
 Ta& 4.090 & 3.73651\\
 V & 4.054 & 4.02202\\
 W & 4.434 & 4.54067\\
\hline
\end{tabular}
\caption{Work functions and number of $d$ electrons 
per unit cell for each metal considered.}
\label{tbl:3}
\end{table}

The behaviour of interface energies versus the difference between work functions  shows a  linear trend as well, 
assigning  the highest interface energies to the smallest work-function differences. 
A rather approximate linear
trend is also observed with respect to differences between the experimental electronegativity values (see Fig.~S2). 
The linearity of both trends in panel (a) of Fig.~\ref{trends-occup}  derives from
work functions and number of $d$ electrons $N^{d}$ being related, as it can be observed in panel (b)  of the same figure.
For the metals here considered, the work function of fcc is larger than that of bcc in all cases. However, the two
sets show approximately linear dependences on the number of $d$ electrons with
slopes of opposite signs (negative for fcc and positive for bcc): the combination of these two conditions
leads to the difference between the two work functions being lower  for higher values of N$^{d}$ 
in both metals (corresponding, in turn, to higher values of their sum, inset of panel (b)). Note that this cannot be extended
to all metals generally. It does not
apply, for instance, to interfaces including Fe (bcc), the work function of which is larger than that of some of
the fcc metals  considered here. 
It is also worth stressing that our calculations do not include magnetic effects, which could be present in
some of the interfaces considered.
For instance, it has been found that the Cr atoms of a flat Cr monolayer buried under a layer of Ag atoms retain a 
significant spin moment {~\cite{das2015}}. Nevertheless, we chose not to include magnetism in our study
in order to compare all interfaces similar conditions, 
avoiding where possible additional effects which might not occur to the same extent in all materials. Magnetism
should, however, be taken into account in a second step to obtain the whole picture, which goes beyond the
scope of the present work. 

\begin{figure}[t]
\begin{center}
\includegraphics[width=\columnwidth]{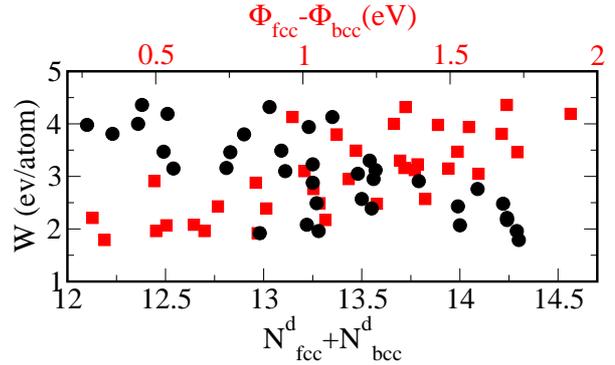}
\caption{Work of separation $W$ as a function of the sum of $d$ electrons (black dots) and difference of work functions (red squares)} \label{adhesion}
\end{center}
\end{figure}

Within the analogy between metal-metal interfaces and metal-vacuum interfaces we also observed (see panel (c)) a dependence
 on the work function for the surface energies, with two distinct linear  trends
for the fcc and the bcc set, the slope of the bcc set being higher.
The bcc crystals generally presented a higher surface
energy than the fcc ones. This is not surprising, given that a higher density of broken bonds
yields to a higher surface energy~\cite{holec2012}.
Within each set, work functions were found to increase for decreasing surface energies, as expected~\cite{wang2014}.
From Table~\ref{interface}, it can be observed that the lowest interface energies in panel (a) correspond to highest work of separation $W$ in panel (b). In fact, a dependence on the sum of the $d$ occupations and on the difference between
work functions was also found for $W$ (Fig. \ref{adhesion}), but with opposite signs.
Our results can be interpreted as follows: when the work functions of two metals are quite different from each other,
charge transfer takes place from fcc to bcc metals with a consequent realignment of their  Fermi levels and formation of interface dipoles.
Overall, the ensuing electrostatic balance contributes to the stabilization
 of the interface, and  this gives rise to a larger work of separation needed
to reestablish the initial situation than if the two work functions were initially closer in value.

\begin{figure}[t]
\begin{center}
\includegraphics[width=\columnwidth]{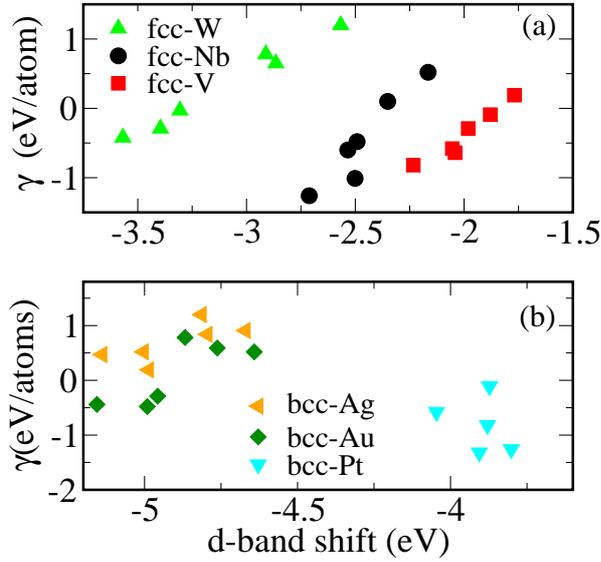}
\caption{Interface energy as a function of the energy distance of the center of mass of the $d$ band from
the common Fermi level in the W, Nb and V atoms at the interface when combined with each of the six fcc metals
 considered (a) and in the Ag, Au and Pt atoms at the interface when combined with each of the six bcc metals considered (b).} \label{dband}
\end{center}
\end{figure}

Inspired by Norskov's $d$-band model \cite{norskov2011}, according to which the bond between a molecule and a metal depends 
on the position of the $d$ band with respect to the Fermi level, we analyzed the position
 of the LDA Kohn-Sham $d$ band (using pseudoatomic projection)  in the atom of the
bcc crystal at the interface when approaching each of the fcc metals considered.
 This  was evaluated as the energetic position of the center of mass of the occupied part of the  band
  with respect to the  global Fermi energy of each interface. For this analysis, we used structures
consisting of 13 layers on each side of the interface.
Interface energies as a function of the so-calculated $d$-band shifts are shown in Fig.~\ref{dband} for W, Nb and V
in panel (a) as examples. An approximately linear behaviour is visible for them all, with the position of the bcc
$d$ band center of mass moving down in energy as the work function of the fcc metal counterpart increases.
 No clear trend was  observed in the opposite case
 (see panel (b) of the same figure, with Au, Ag and Pt as examples), i.e., 
referring to the energetic position of the $d$ band in the fcc atom at the
 interface with each of the six bcc crystals.

\begin{figure}[t]
\begin{center} 
\includegraphics[width=\columnwidth]{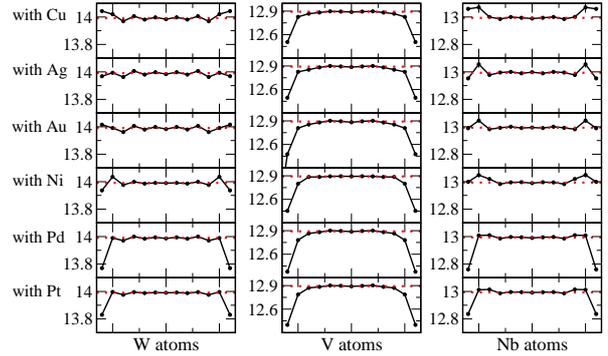}
\caption{L\"{o}wdin charges for all atoms in W, V and Nb when interfaced
with each of the six fcc metals considered. The horizontal red dotted line indicates the bulk value.} \label{charges}
\end{center}
\end{figure}

\begin{figure}[t]
\begin{center} 
\includegraphics[width=\columnwidth]{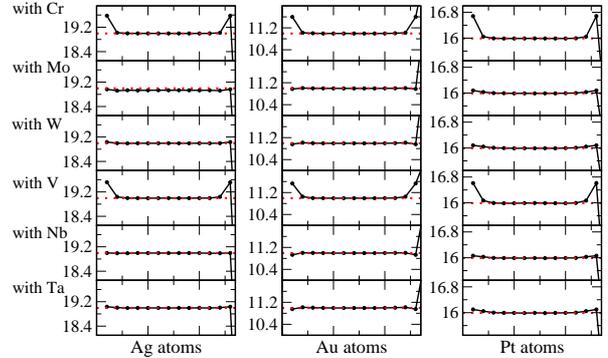}
\caption{L\"{o}wdin charges for all atoms in Ag, Au and Pt when interfaced
with each of the six bcc metals considered. The horizontal red dotted line indicates the bulk value.} \label{chargesfcc}
\end{center}
\end{figure}

For further insight, we analyzed the pseudoatomic L\"{o}wdin charges across the junctions for the same systems as in Fig.~\ref{dband}.
Such charges are shown for the W, V and Nb atoms  in Fig.~\ref{charges}. 
In all cases, the population at the bcc atom at the interface decreases as the work function of the fcc metal increases
 (as expected with increasingly favourable conditions for charge transfer from bcc to fcc).
The same kind of analysis does not show, however, a clear trend for the fcc metals (see Fig.~\ref{chargesfcc};
for Pt, the complete charge profile across the whole junction is shown in Fig.~S3 of the Supporting Information).
As an aside, we  observed a general tendency of the bcc metal to present charge oscillations across the junction more than in
the fcc metal.   

%
%

%
\begin{figure}[t]
\begin{center}
\includegraphics[width=\columnwidth]{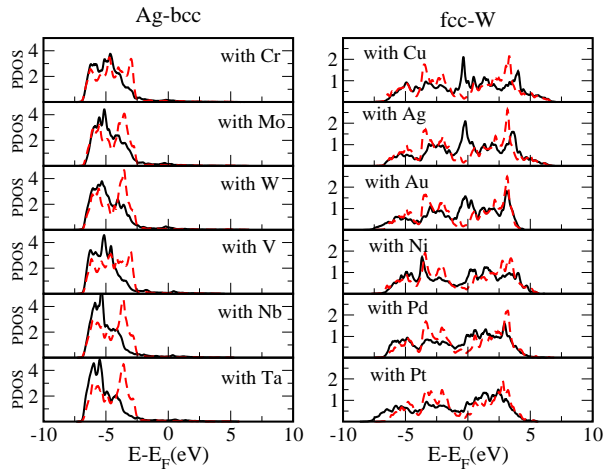}
\caption{Projected density of state (PDOS)  in the atom at the interface (solid black line) and in a inner layer (red dashed
line) in Ag (left) and W (right) when combined
with each of the bcc (fcc) metals considered. All energetics position are evaluated with respect to the global Fermi
level of the interface.} \label{dos-layers}
\end{center}
\end{figure}

Overall, our calculated charge distributions show that only the atoms in one or two layers close to the  interface are strongly affected by the
contact with the other metal. In Fig.~\ref{dos-layers} we compare, in Ag (fcc) and W (bcc) as examples, the profile
of the $d$ band in the atom at the very interface (solid black line) with that in an atom in the middle of the same block (dashed red line).
It can be observed that
the profile of the $d$ band in the inner layer approximately stays constant regardless the metal Ag (W) is combined with;
conversely, it changes quite
drastically at the interface, depending on the metal at the other side.
Finally, we noticed a tendency of the fcc crystal to modify its  lattice parameter in the plane parallel to
the interface (with a$_\textrm{fcc}/\sqrt{2}$ increasing by up to 0.7~\AA, see Fig. S4 of the Supporting Information) more than in bcc
 (for which a$_\textrm{bcc}$ tends to
 remain unaffected, with an average variation of 0.01~\AA).

\section{Conclusions} \label{sec-conclusions}
In summary, we have demonstrated a novel way of computing interface energies. Our method was constructed as an extension of  the
Fiorentini method, which was developed for calculating surface energies.
We studied fcc-bcc metal-metal interfaces in the (100) orientation. We found a dependence of the interface energy
 on the difference between the work functions of the two metals as well as the total number of $d$ electrons per unit cell.
We believe that such relationships between interface energies and bulk properties can be useful for selecting
 materials and optimizing their adhesion.
Interestingly, the trend of the interface energy is reflected in the position of the $d$ band in the bcc metal 
with respect to the common interface Fermi level, whereas the same does not apply for its fcc counterpart. 
For the systems with the lowest interface energies, L\"{o}wdin charges show a clear sign of charge depletion in
the bcc metal  at the boundaries. 
The reliable first-principles prediction of surface and interface energetics is a long-standing challenge of
 significant technological relevance. The efficient methodology proposed and trends revealed in this work now provide
 a new route to achieving this for metals. It remains to be seen whether the energetic effects of complicating factors
 such as strain and reorganisation, chemical bonding effects such as at oxide interfaces, and the emergence of interface
 magnetization and polarization are also amenable at least semi-quantitatively to simple bulk descriptors. This is a
 promising avenue for future investigation.

\section*{Acknowledgements}
This publication has culminated from research supported in part by a research grant from Science Foundation
 Ireland (SFI) under Grant Number SFI/12/RC/2278. LAZ acknowledges support from  the Spanish MINECO through the grant
 MAT2014-58982-JIN. 
All calculations were performed on the Boyle cluster maintained by the Trinity Centre for High Performance
 Computing. This cluster was funded by the European Research Council under the Quest project.
The authors would like to thank John Donegan and David McCloskey of Trinity College Dublin for helpful discussions.


\section*{References}

\end{document}